# Nowcasting earthquakes with QuakeGPT off Hokkaido and Tohoku, Japan

**Running title:** Earthquake nowcasting off Hokkaido–Tohoku

**Authors:** K. Z. Nanjo[1,2,3,4,5]*, J. Yazbeck[5], I. T. Baughman[5], J. Donet[5], and J. B. Rundle[5,6,7]

[1] Global Center for Asian and Regional Research, University of Shizuoka, Shizuoka, Japan

[2] Center for Integrated Research and Education of Natural Hazards, Shizuoka University, Shizuoka, Japan

[3] Institute of Statistical Mathematics, Tokyo, Japan

[4] Japan Agency for Marine-Earth Science and Technology, Yokohama Institute for Earth Sciences, Yokohama, Japan

[5] Department of Physics and Astronomy, University of California, Davis, CA, USA

[6] Department of Earth and Planetary Sciences, University of California, Davis, CA, USA

[7] Santa Fe Institute, Santa Fe, NM, USA

***Correspondence:** K. Z. Nanjo, Global Center for Asian and Regional Research, University of Shizuoka, Shizuoka, Japan. Email: nanjo@u-shizuoka-ken.ac.jp

**ORCID:** K. Z. Nanjo, https://orcid.org/0000-0003-2867-9185; J. Yazbeck, https://orcid.org/0000-0003-0223-5260; I. T. Baughman, https://orcid.org/0009-0005-7895-3084; J. B. Rundle, https://orcid.org/0000-0002-1966-4144

**Funding information:** Ministry of Education, Culture, Sports, Science and Technology (MEXT) of Japan, Second Earthquake and Volcano Hazards Observation and Research Program (Earthquake and Volcano Hazard Reduction Research) and Earthquake Disaster Prevention Research Project to Minimize Damage from Gigantic Earthquakes such as the Nankai Trough Earthquake and Contribute to Rapid Recovery and Reconstruction; SECOM Science and Technology Foundation, FY2023 Special Research Grant; Southern California

Earthquake Center, Grant SCON-00007927; John LaBrecque Fund, University of California, Davis.

## Abstract

Large earthquakes in subduction zones cause damage and tsunamis. Preparedness requires knowing how the potential for the next large earthquake changes. Earthquake nowcasting tracks this change with an indicator, the earthquake potential state (EPS), computed from monthly numbers of small earthquakes. Here we further developed earthquake nowcasting in two ways and applied it to the 2013–2026 Japan Meteorological Agency catalogue of the subduction zone off Hokkaido and Tohoku, Japan. First, instead of selecting the single best EPS curve, we averaged many EPS curves, providing greater weight to better curves. In the year before a large earthquake (magnitude 6.5 or larger), this ensemble EPS tended to indicate a higher potential than at other times. Random numbers used in place of the ensemble EPS did not show this tendency. Second, we trained QuakeGPT, a neural network, on earthquakes simulated for the region and used it to compute the ensemble EPS for the following month and beyond the last month of data. For the following month, QuakeGPT computed values that were almost the same as those of the observed ensemble EPS, except when a new period of many earthquakes began. Beyond the data, QuakeGPT continued the ensemble EPS, with a range, for about one year. The ensemble EPS and its continuation by QuakeGPT can be updated monthly from existing earthquake monitoring, so they can be used together with long-term assessments and with information issued after large earthquakes, as one part of multi-hazard monitoring for preparedness in subduction zones.

# 1 Introduction

Losses from large earthquakes can be reduced by decisions taken before the earthquakes. Examples include strengthening buildings, planning tsunami evacuation, storing supplies, and assigning resources for the response. A large earthquake in a subduction zone induces strong shaking, a tsunami, and a long period of active seismicity. Decisions therefore must consider several hazards that follow one another, i.e., cascading hazards. To take these decisions, information on earthquake risk is needed. The information should be quantitative, easy to interpret, and updated as new observations emerge. The information should also be derived from existent monitoring.

Two kinds of information are available. Long-term assessments give the probability of large earthquakes over decades (Field et al., 2014; Headquarters for Earthquake Research Promotion [HERP], 2017, 2019). Although these assessments are quantitative and easy to interpret, they change slowly. Operational forecasts are issued after large earthquakes (Jordan et al., 2011). In Japan, the Japan Meteorological Agency (JMA) issues an outlook of seismic activity for the days to weeks following a large earthquake (HERP, 2016), with probabilities estimated from a statistical model of aftershocks (HERP, 1998). Between the two assessment options, an indicator lasting months to years is missing. Such an indicator would follow how the potential for large earthquakes changes, using routine earthquake monitoring. This indicator would also fill in the gap between the slowly changing assessments and the short-lived forecasts, and would connect monitoring to preparedness.

The term nowcasting emerged from the fields of economics and finance, where it means estimating the uncertain current state of a system by indirect means (Rundle et al., 2016). Earthquake nowcasting applies this idea to earthquakes and was developed to provide such an indicator from routine earthquake monitoring (Rundle et al., 2016, 2023, 2024, 2026). Nowcasting has used several indicators. As one example, it has been based on the number of small earthquakes since the last large earthquake (Rundle et al., 2016, 2026). Nowcasting has been applied to California (Rundle et al., 2022, 2023, 2024), to cities around the world, including Tokyo (Rundle et al., 2016), and to Greece, the Himalayas, Iran, and the Bay of Bengal (Pasari, 2019, 2023; Pasari & Sharma, 2020; Chouliaras et al., 2023).

The indicator used in this study is the earthquake potential state (EPS) of Rundle et al. (2025). The EPS is the logarithm of an exponential moving average (EMA) of the monthly numbers of small earthquakes. When plotted on an inverted axis, the EPS decreases in a month with many earthquakes and increases in the quiet months that follow (Rundle & Donnellan, 2020; Rundle et al., 2022, 2025). In California, large earthquakes tend to occur in these quiet months (Rundle et al., 2022, 2023, 2025).

Nowcasting describes the present state (Rundle et al., 2016). To compute future values of the EPS, Rundle et al. (2025) introduced QuakeGPT, a transformer (Vaswani et al., 2017), which is a type of neural network. Earthquake catalogues (lists of recorded earthquakes) are too short for training, so QuakeGPT was trained and tested on synthetic catalogues, that is, catalogues produced by computer simulation. Rundle et al. (2025) also extended the EPS with QuakeGPT beyond the end of the catalogue used for testing, where there are no values of the EPS to compare with. Importantly, Rundle et al. (2025) did not apply QuakeGPT to an observed catalogue. Other neural-network models, including transformers, have also been applied to earthquake nowcasting (Fox et al., 2022; Jafari et al., 2024).

The Pacific plate boundary off Hokkaido and Tohoku is a subduction zone with highly active seismicity (Figure 1). This plate boundary produced the 2011 Tohoku earthquake of magnitude ($M$) 9.0. Great earthquakes along this plate boundary remain a central concern in the preparedness for tsunamis and cascading hazards in Japan (HERP, 2017, 2019; Cabinet Office, 2021). Japanese agencies operate dense observation networks on land, and on the seafloor (Obara et al., 2005; Aoi et al., 2020). The region is therefore well suited to examining

EPS nowcasting in a subduction zone. This setting differs from California, where the EPS and QuakeGPT were developed (Rundle et al., 2022, 2025).

To our knowledge, earthquake nowcasting has not yet been applied to a region with seismicity as active as that off Hokkaido and Tohoku. It is also not known whether the EPS and QuakeGPT can be considered as a basis for providing information for preparedness, and if so, how and with what limits.

This study asks two questions. Does EPS nowcasting work in a subduction zone? Can QuakeGPT, trained on simulated earthquakes, compute the EPS for the next month and beyond the data period? For the first question, we applied nowcasting to the JMA earthquake catalogue off Hokkaido and Tohoku from 2013 onward (Figure 1). For the second question, we trained QuakeGPT on a 1000-year synthetic catalogue for this region, simulated with a statistical model of earthquake occurrence. We tested QuakeGPT on years of this catalogue that were not used for training and on the EPS of the observed catalogue (Section 3.3). Beyond the data, we started QuakeGPT from past months of the observed catalogue and compared the computed values with the EPS of later months (Section 3.3). In Section 5, we discuss what the EPS measured, what QuakeGPT learnt, and what the limits of both are. We also discuss what this study can contribute to the multi-hazard monitoring needed for societal resilience in a subduction zone (Section 5.2).

## 2 Data

The Pacific plate subducts beneath northeastern Japan at the trench axis (Figure 1). The 2011 $M$9.0 Tohoku earthquake (Ozawa et al., 2011; Simons et al., 2011; Nanjo et al., 2012b) and the activity that followed it dominated the earthquake activity of the region for several years.

We used the JMA earthquake catalogue, and downloaded it on 10 October 2026. The analysis covers the period from 1 January 2013 to 8 October 2026, which we refer to as the study period. We started the analysis in 2013, so that the EPS was not dominated by the decay of activity following the 2011 earthquake. JMA compiles the catalogue from its own seismograph network. Since 1997, JMA has also used, in real time, seismic records from seismograph networks operated by Japanese universities and institutions (Obara et al., 2005; Nanjo et al., 2010). In September 2020, JMA added the records of a seafloor observation network off the Pacific coast (Aoi et al., 2020). This addition made the depths of earthquakes far from the coast more accurate. Prior to that date, there had been few offshore stations. Earthquakes more than about 100 km from the coast were therefore located over a wide range of depths. Nanjo et al. (2026) assumed that these earthquakes occurred around the plate boundary. A comparison with an independent earthquake catalogue supported this assumption (Nanjo et al., 2026).

The study region (Figure 1) and the selection of earthquakes follow Nanjo et al. (2026). Earthquakes on the seaward side of the trench are excluded. Depths are limited to 120 km or less. Within 100 km of the coast, we maintained the earthquakes within 20 km above or below the plate boundary (Nakajima & Hasegawa, 2006; Kita et al., 2010). Beyond 100 km from the coast, the depths before September 2020 are unreliable. At that distance, we maintained all earthquakes, following the assumption of Nanjo et al. (2026). We used $M$ 2.5 as the lower magnitude throughout. Nanjo et al. (2026) discarded the earthquakes below this magnitude, allowing the catalogue to be uniform over that period. In the study period, the region experienced a mean of 282 earthquakes with $M \geq 2.5$ per month of about four weeks (Text S1).

Nowcasting requires a set of target earthquakes. The target earthquakes included the large earthquakes whose potential the EPS is meant to indicate, and the skill of the EPS was evaluated against them (Section 3.2). We adopted earthquakes with $M \geq 6.5$ as targets. In this region, earthquakes of this size can cause damage and tsunamis. Targets with lower and higher $M$ are examined in Section 4.2. Nineteen target earthquakes with $M \geq 6.5$ occurred between 2013 and October 2026 (Figure 1). Seven of them had $M \geq 7.0$. Considering that the evaluation was made monthly (Section 3.2), targets that occurred within the same month were counted once (17 and 20 February 2015; two on 9 November 2025; 8, 9, and 12 December 2025; 26 March and 20 April 2026).

## 3 Methods

We used the methods of Rundle et al. (2025). Section 3.1 summarises them. Sections 3.2 and 3.3 state where this study differs from those methods, in terms of the nowcast and QuakeGPT, respectively. The details, including the equations, are provided in Texts S1–S5.

### 3.1 Methods of Rundle et al. (2025)

The first method is the nowcast. The earthquakes above a lower magnitude are counted in each month. An EMA of the monthly numbers is used, because it provides greater weight to recent months. The EPS is the logarithm of this average. Θ represents a single EPS curve. The formula for Θ is Equation (S3) in Text S1. The axis for Θ is inverted, like the vertical axis of Figure 2. Following Rundle et al. (2025), Θ is described as it appears on this axis. For Θ, words such as high, low, above, and below refer to positions on this axis, and larger and smaller to its values. On this axis, Θ decreases in a month with many earthquakes, as after a large earthquake. In the quiet months that follow, Θ increases and returns to the level set by the floor, where this floor is the mean monthly number (Section 2) multiplied by a factor $\lambda_{min}$, the floor parameter. A higher Θ indicates a higher potential for a large earthquake.

Rundle et al. (2025) evaluated Θ with the receiver operating characteristic (ROC; Brown & Davis, 2006). Each month is labelled based on whether a target earthquake occurs within a future time window, $T_W$, after the end of the month. The area under the ROC curve (AUC;

Hanley & McNeil, 1982) measures how well a high Θ separates the months followed by a target earthquake within $T_W$ from the other months. Rundle et al. (2025) selected the floor and averaging length $L$ (the number of recent months that carry most of the weight in the EMA) that gave the largest AUC. For a region centred on Los Angeles, California, Rundle et al. (2025) used $\lambda_{min}$ = 1.2, $L$ = 36 months, and $T_W$ = 3 years.

The second method is QuakeGPT (Section 1). QuakeGPT computes Θ of the next month from Θ of the preceding 36 months. QuakeGPT was trained on Θ of a long synthetic catalogue, which Rundle et al. (2025) produced with a stochastic simulation model (Rundle et al., 2024). To continue Θ further ahead, the values computed by QuakeGPT are used as new input, month after month.

## 3.2 Differences in this study: the nowcast

Four settings differ from those employed by Rundle et al. (2025). First, the averaging length is $L$ = 20 months. Second, the future time window is $T_W$ = 1 year. Third, we applied no cap. The code published by Rundle et al. (2025) allowed a cap, an upper limit on the monthly numbers set by a cap parameter, $\lambda_{max}$. Fourth, we did not evaluate the first year, 2013, which we call the start-up year (grey area in Figure 2). The EMA started in January 2013 without earlier months, for the reason given in Section 2, and continued to build up in that year. Section 4.2 shows the effects of the first three settings (Figures S1 and S2 for $L$ and the cap).

The main difference from Rundle et al. (2025) is that we did not select the single EPS curve with the largest AUC (Section 3.1). Instead, we adopted a weighted mean of EPS curves computed with many values of the floor, so that the result does not depend on one choice of the floor. We used 101 values of $\lambda_{min}$, from 1.00 to 6.00 in 0.05 steps. The 101 EPS curves are referred to as members. Thus, $\Theta_j$ represents member $j$ ($j$ = 1, …, 101). Each member was weighted according to its own AUC, $AUC_j$. A member with a larger AUC has a larger weight, and a member with an AUC below the mean of all members has no weight. We call these weights z-score weights, where the weight is similar to that used by Holliday et al. (2005) and Nanjo et al. (2006a, 2006b) (Figure S3). The weighted mean of the members is the ensemble EPS, $\Psi$. As with $\Theta$ (Section 3.1), words such as high, low, above, and below refer to positions of $\Psi$ on the inverted axis, and larger and smaller to its values. In the figures, the smallest value of $\Psi$, $\Psi_{min}$, is its highest position (dashed line in Figure 2). The bands around $\Psi$ in Figure 2 contain the central 50% and 80% of the members, after each member is shifted so that the level set by its floor lies at $\Psi_{min}$ (Text S2 and Figure S4). The weights were set with the target earthquakes of the observed catalogue, which are also used for the evaluation. Details are provided in Text S2.

The evaluation follows Rundle et al. (2025). A month is referred to as a positive month if at least one target earthquake occurs within $T_W = 1$ year after its end, and as a negative month otherwise. We evaluated the 153 months that ended between January 2014 and October 2025, which is the last month that was followed by one full year of data. The fraction of positive months among them, 0.51, is called the base rate. The skill of $\Theta$ or $\Psi$ refers to how well a high value separates the positive months from the negative months. For each threshold, an alarm is declared in the months in which $\Theta$ or $\Psi$ lies above the threshold in the figures, that is, in which $\Theta$ or $\Psi$ is smaller than the threshold. The hit rate is the fraction of positive months with an alarm, and the false-alarm rate is the fraction of negative months with an alarm. The ROC curve plots the hit rate against the false-alarm rate as the threshold is varied (Figure 3a). The AUC is 0.5 without skill and 1 for a perfect separation.

A second measure of skill, the average precision (AP), was added to the evaluation of Rundle et al. (2025). AP is the area under the precision–recall curve (Figure 3b), which is more informative than the ROC curve when there are few positive months (Saito & Rehmsmeier, 2015), as they are for short $T_W$ (Section 4.2). Precision is the fraction of alarm months that are positive, and recall is another term for the hit rate. Without skill, AP equals the base rate (Saito & Rehmsmeier, 2015). Rundle et al. (2025) compared the ROC curve of Θ with that of 200 random time series. We compared the ROC and precision–recall curves of Ψ with those of 200 series of random scores (Figure 3). The details of the evaluation are provided in Text S3.

### 3.3 Differences in this study: QuakeGPT

Rundle et al. (2025) trained QuakeGPT on a long synthetic catalogue from their own simulation model and tested QuakeGPT on a second synthetic catalogue from the same model. We instead simulated one synthetic catalogue of 1000 years for the study region, which we call the 1000-year catalogue (Figure S5). We divided this catalogue in time (Figure 4). Years 11–800 are the training data, years 801–900 are the choosing data, and years 901–1000 are the test data. The choosing data were used to select the QuakeGPT that was kept after the training. Ψ of the first 10 years was not used. These years are the start-up of the EMA, like 2013 for the observed catalogue (Section 3.2). The catalogue was long enough to leave out 10 years.

The catalogue was simulated with the temporal epidemic-type aftershock sequence (ETAS) model (Ogata, 1988, 2006; Ogata & Tsuruoka, 2016). In this model, the rate of earthquakes is a constant background rate plus a rate added by each earlier earthquake (Text S4). The ETAS model was fitted to the earthquakes of 2023–2026 in and around the study region (Figure 1). The simulated earthquakes follow the Gutenberg–Richter frequency–magnitude relation (Gutenberg & Richter, 1944), in which a constant $b$ describes the relative numbers of large and small earthquakes (Text S4). We set $b = 0.85$, a value that the observed catalogue follows (Figure 5). The largest magnitude was set to $M_{\max} = 9.0$, the $M$ of the 2011 Tohoku earthquake. In Figure 5, the 1000-year catalogue (black) and the observed catalogue (blue) are close from $M$ 2.5 to $M$ 7.7. Above about $M$ 8.5, the 1000-year catalogue falls below the Gutenberg–Richter relation (red line), because $M$ is limited to $M_{\max}$. We computed Ψ of the 1000-year catalogue with the same 101 values of $\lambda_{min}$ and the same weights as $\Psi$ of the observed catalogue, so that $\Psi$ was defined in the same way in the training and tests (Text S4).

We took QuakeGPT from the code published by Rundle et al. (2025), which we call the published code, and trained QuakeGPT as shown in Figure 4. We did not change the structure of QuakeGPT. QuakeGPT computed $\Psi$ of a month from $\Psi$ of the 36 months before it, which we call the input window. $\Psi_{QuakeGPT}$ indicates the value of $\Psi$ computed by QuakeGPT. Each input window was rescaled so that its smallest value becomes 0 and its largest value 1. This rescaling removed the difference in the values of $\Psi$ between the 1000-year catalogue and the region (Figures 2 and 6). QuakeGPT could thus be trained on the 1000-year catalogue and applied to the region (Text S5). During training, QuakeGPT was adjusted so that $\Psi_{QuakeGPT}$ came closer to the true values, i.e., to the values of $\Psi$ known from the catalogue. The training was performed in passes. In one pass, QuakeGPT learned once from every input window of the training data. We completed 200 passes. After each pass, QuakeGPT computed $\Psi_{QuakeGPT}$ for each month of the choosing data. These values were compared with the true values by the root-mean-square error (RMSE), the square root of the mean squared difference between $\Psi_{QuakeGPT}$ and the true values. The QuakeGPT of the pass with the smallest RMSE was kept. Figure 6 shows $\Psi_{QuakeGPT}$ after 1, 10, and 200 passes, representing the effect of the training.

Two changes to the published code were needed for a valid test (thick boxes in Figure 4). First, $\Psi_{QuakeGPT}$ of a month was computed only from $\Psi$ of the 36 months before it. In the published code, $\Psi_{QuakeGPT}$ of a month could also use the true values of that month and of later months. Second, we selected the QuakeGPT that is kept on the choosing data. In the published code, the test data were the choosing data: that is, the test data were also used in the choice (the QuakeGPT was kept on the test data). Further changes to the published code are described in Text S5.

QuakeGPT was tested on values of $\Psi$ that were not used for training nor for choosing. These values of $\Psi$ are the test data of the 1000-year catalogue (Figure 6) and the observed $\Psi$, i.e., $\Psi$ of the observed catalogue (Figures 2 and 7). QuakeGPT was used in two ways, which we refer to as "the next month" and "beyond the data." For the next month, QuakeGPT computed $\Psi_{QuakeGPT}$ of a month from the true values of $\Psi$ in the preceding 36 months. Beyond the data, as in Rundle et al. (2025), QuakeGPT was provided the true values of $\Psi$ up to a month that we call the start month. After the start month, $\Psi_{QuakeGPT}$ was used as new input, month after month.

We compared $\Psi_{QuakeGPT}$ with $\Psi$ and with persistence. Persistence assumes that $\Psi$ keeps its last value. Persistence needs no training, so the comparison with persistence shows what the training adds. For the next month, persistence is the $\Psi$ of the prior month. Beyond the data, persistence is the $\Psi$ of the start month (Figures 7a, b). The relative intensity model makes a similar assumption in space (Holliday et al., 2005; Nanjo et al., 2006a, 2006b; Nanjo, 2010, 2011). This model assumes that future large earthquakes are likely to occur where small earthquakes were frequent in the past. To test $\Psi_{QuakeGPT}$ beyond the data, we used as start months all months from October 2015, the first month that completes an input window of 36 months, to September 2026, the last month that is followed by another month of $\Psi$. For each start month, $\Psi_{QuakeGPT}$ of each month after the start was compared with $\Psi$ of the same month, by the absolute difference $|\Psi - \Psi_{QuakeGPT}|$ (Figure 7). Details are provided in Text S5.

# 4 Results

## 4.1 The ensemble EPS, Ψ

Figure 2 shows $\Psi$ from 2013 to October 2026, computed with the settings defined in Section 3.2. The black line and bands show four things. First, $\Psi$ changes in three ways only. $\Psi$ stayed at $\Psi_{min}$, decreased abruptly in the months of some large earthquakes, and returned to $\Psi_{min}$ in about three years. Second, $\Psi$ decreased by more than 0.01 three times, for the *M*7.4 earthquake of November 2016, the *M*7.3 earthquake of February 2021, and the *M*7.4 earthquake of March 2022. Third, the size of a decrease in $\Psi$ was not determined by the *M* of the earthquake. The largest decrease came with the *M*7.4 earthquake of November 2016, whereas the *M*7.7 earthquake of April 2026, the largest of the study period, hardly decreased $\Psi$. Fourth, the bands around $\Psi$ had no width while $\Psi$ was at $\Psi_{min}$. The bands became narrower as $\Psi$ returned to $\Psi_{min}$.

Figure 3 shows that $\Psi$ tends to be higher in the positive months than in the negative months. The curve of $\Psi$ (blue) lies far above the dashed line and above almost all the 200 curves of random scores. The values of the areas under the two blue curves are AUC = 0.870 and AP = 0.922 (base rate 0.51).

## 4.2 Dependence on settings

This section shows how the skill of $\Psi$ (Section 4.1) and of $\Theta$ depends on $L$, the cap, $T_W$, and $M$ of the target earthquakes. Figures S1, S2, and S6 show five things. First, $L$ = 20 months is a typical value (Figure S1). Figure S1 shows an AUC of $\Theta$ against $L$ for $\lambda_{min}$ = 2.5 and 4.0. For both values of $\lambda_{min}$, the AUC is close to 0.5 at $L$ = 1 month and reaches about 0.85 at about $L$ = 5 months. At a longer $L$, the AUC changes only slowly. The grey areas mark the typical values of $L$ defined in Text S1, and $L$ = 20 months lies in both grey areas. Rundle et al. (2025) used $L$ = 36 months, which lies in the grey area for $\lambda_{min}$ = 4.0 only (vertical dotted line in Figure S1).

Second, for both values of $\lambda_{min}$ in Figure S1, a cap did not increase the AUC at $L$ = 20 months. We compared the AUC for four values of $\lambda_{max}$, 1, 3, 5, and $\infty$, where $\infty$ indicates no cap (Text S1). Leaving out the cap also removes one adjustable parameter.

Third, $L$ = 10 months is also a typical value (blue circles in Figure S1). With $L$ = 10 months, the ROC curve of $\Psi$ is almost the same as with $L$ = 20 months (Figure S2b). With $L$ =

10 months, Ψ decreased in the same months as with $L = 20$ months. The decreases are larger, and Ψ returns to $\Psi_{min}$ sooner (Figure S2a).

Fourth, for targets of $M \geq 6.5$, at $T_W = 0.25$, 0.5, 1, and 2 years, the AUC of Ψ is larger than 0.5 and the AP is larger than the base rate (black in Figures S6b and S6c). A longer $T_W$ makes more months positive. The base rate is about 0.2 at $T_W = 0.25$ years and about 0.65 at $T_W = 2$ years (Figure S6a). We used $T_W = 1$ year, because at this value, there is an almost equal number of positive and negative months (Figure S6a).

Fifth, with targets of $M \geq 6.0$ or $M \geq 7.0$ (blue and red, respectively in Figure S6), every AUC shown is also larger than 0.5 (Figure S6b). Every AP shown for these targets is larger than the base rate (Figure S6c). With targets of $M \geq 6.0$, almost every month is positive at $T_W = 1$ year, and every month is positive at $T_W = 2$ years (Figure S6a).

### 4.3 QuakeGPT for the next month

Figure 6 shows Ψ in the test data of the 1000-year catalogue, with $\Psi_{QuakeGPT}$ for the next month and with persistence (crosses in Figure 6b). This figure shows three things. First, after 200 passes, $\Psi_{QuakeGPT}$ is almost the same as Ψ (Figure 6a). In the month of the *M*9.0 earthquake, however, Ψ decreases, while both $\Psi_{QuakeGPT}$ and persistence retain the value of the previous month, far from Ψ (Figure 6b). Second, $\Psi_{QuakeGPT}$ approaches Ψ with more passes (pink, purple, and green dots in Figure 6b). Third, persistence is also close to Ψ. Persistence follows Ψ one month late, which can be seen after a decrease of Ψ (Figure 6b).

Figure 2 shows the observed Ψ with $\Psi_{QuakeGPT}$ for the next month (green dots) and with persistence (crosses). The green dots and crosses show three things. First, as in the 1000-year catalogue (Figure 6a), $\Psi_{QuakeGPT}$ for the next month is almost the same as Ψ (Figure 2). Second, as in the 1000-year catalogue (Figure 6b), persistence is also close to Ψ and follows Ψ one month late (Figure 2). Third, as in the 1000-year catalogue (Figure 6b), in the months in which $\Psi$ decreases abruptly, both $\Psi_{QuakeGPT}$ and persistence stay at the value of the month before (Figure 2).

## 4.4 QuakeGPT beyond the data

Figures 7a and 7b show two examples of $\Psi_{QuakeGPT}$ beyond the data, which started in January 2017 and May 2022, with persistence (crosses). These panels show three things. First, in Figure 7a, $\Psi_{QuakeGPT}$ is almost the same as $\Psi$ for more than a year and then departs from $\Psi$ gradually. Second, in Figure 7b, $\Psi_{QuakeGPT}$ departs from $\Psi$ within the first months. Third, persistence keeps the $\Psi$ of the start month, whereas $\Psi$ returns to $\Psi_{min}$, so that persistence departs from $\Psi$ from the first month on (Figures 7a, b). From here on, we only consider QuakeGPT.

Figures 7c and 7d show $|\Psi - \Psi_{QuakeGPT}|$ for all start months (Section 3.3). These panels show three things. First, the median of $|\Psi - \Psi_{QuakeGPT}|$ grows mainly during the first year after the start, to about 0.003. Later, the median stays between about 0.003 and 0.004 (Figure 7d). Second, the spread of $|\Psi - \Psi_{QuakeGPT}|$ is large compared with its median. For example, 13 months after the start, the median is about 0.003, whereas the central 95% reaches about 0.018 (Figure 7d and the inset of Figure 7c). Third, for the example of Figure 7a, $|\Psi - \Psi_{QuakeGPT}|$ lies close to the median in the first months and below the median afterwards (green dots in Figure 7d). For the example of Figure 7b, $|\Psi - \Psi_{QuakeGPT}|$ lies above the median in every month (green triangles in Figure 7d).

For the year after the last month of data, QuakeGPT gives a path of $\Psi$ (green open circles in Figure 2), with a range from the start months in the past (green dotted lines). The dotted lines lie at the median of $|\Psi - \Psi_{QuakeGPT}|$ above and below the open circles (black line in Figure 7d). For half of the start months in the past, $\Psi_{QuakeGPT}$ differed from $\Psi$ by less than this median (solid line at a fraction of 0.5 in Figure 7c). We show $\Psi_{QuakeGPT}$ beyond the data in Figure 2 for the first year only, the time in which the median grew most (Figure 7d).

## 5 Discussion

### 5.1 What the EPS measures off Hokkaido and Tohoku

$\Psi$ decreases in a month with many earthquakes. In the quiet months that follow, $\Psi$ returns to $\Psi_{min}$, at a pace set by $L$ of the EMA (Sections 3.1, 4.1, and 4.2). The size of a decrease thus shows how many earthquakes the last period of activity had. How far $\Psi$ has returned to $\Psi_{min}$ since then shows how long ago that period was. $\Psi$ is therefore high in quiet months long after the last period of activity, or after a period of activity with few earthquakes.

As Figure S3a shows, $AUC_j$ depends on $\lambda_{min}$, whereas $\Psi$ needs no choice of one value of $\lambda_{min}$. Rundle et al. (2025) selected the EPS curve with the largest AUC (Section 3.1). Among our members, the one with the largest $AUC_j$ is marked in red in Figure S3. Just above the $\lambda_{min}$ of this member, $AUC_j$ is much smaller, and from $\lambda_{min}$ of about 5 on, $AUC_j$ is 0.5, the value without skill (Figure S3a). The member with the largest $AUC_j$ clearly decreased only with the earthquake of November 2016 (purple curve in Figure S4b). This member decreased only slightly with the earthquakes of February 2021 and March 2022, and not at all with the activity of 2025–2026 (Figure S4b). The large $AUC_j$ of this member therefore depends on which single months exceed its floor. $\Psi$ is the weighted mean of the members over all 101 values of $\lambda_{min}$ (Section 3.2), and $\Psi$ decreases with each of these periods of activity (Figure 2).

### 5.2 What did QuakeGPT learn?

After 200 passes, $\Psi_{QuakeGPT}$ for the next month is almost the same as $\Psi$ in the test data (Figure 6) and in the region (Figure 2). What QuakeGPT has learned is how $\Psi$ returns to $\Psi_{min}$ after a decrease of $\Psi$ (Section 4.3). QuakeGPT thus follows how activity proceeds after it has begun, rather than looking for signs before it. Short-term models of earthquake occurrence have the same aim, for example the Short-Term Earthquake Probability (STEP) model, studied for seismicity in California (Gerstenberger et al., 2005), and the space-time ETAS model (Ogata, 2011), studied for activity after the 2011 Tohoku earthquake (Nanjo et al., 2012a). What QuakeGPT has learned from the catalogue applies also to the region, for two reasons. First, rescaling of the input windows removes the difference in the values of $\Psi$ between the 1000-year catalogue and the region (Section 3.3). Second, in the region, as in the 1000-year catalogue, $\Psi$ returns to $\Psi_{min}$ in the quiet months after a decrease of $\Psi$ (Figures 2 and 6a). However, $\Psi_{QuakeGPT}$ does not decrease in the month in which new activity begins (Section 4.3).

Beyond the data, QuakeGPT gives a path of $\Psi$ for about one year, with a range from the start months in the past (Figures 2 and 7). The path consists of values of $\Psi$, not of earthquakes. The present $\Psi$ already concerns target earthquakes in the year after the last month of data, because the evaluation of $\Psi$ asks, for each month, whether a target earthquake occurs within one year after the end of the month (Section 3.2; Figure 3). The present $\Psi$ and the path of $\Psi$ beyond the data can both be updated every month from earthquake monitoring that already exists (Section 2). For preparedness, the present $\Psi$ and the path can therefore be used together with long-term assessments and with the information issued after large earthquakes (Section 1). The present $\Psi$ and the path can thus be one part of the multi-hazard monitoring of a subduction zone.

This study differs from Rundle et al. (2025) in how QuakeGPT was tested (thick boxes in Figure 4). In our tests, $\Psi_{QuakeGPT}$ for a month was unable to draw on the true value of $\Psi$ in that month (Section 3.3). The test data also play no part in the choice of the QuakeGPT that was kept. In addition, QuakeGPT was tested on the observed $\Psi$ (Figures 2 and 7). $\Psi_{QuakeGPT}$ beyond the data was compared with $\Psi$ from every start month in the past (Figure 7). Rundle et al. (2025) showed such values only after the end of their second synthetic catalogue, where no true values exist.

### 5.3 Limitations

The weights of the members were set with the same target earthquakes that were used for the evaluation (Section 3.2). Therefore, the AUC of $\Psi$ may be lower for target earthquakes that were not used to set the weights. An evaluation of the months after 8 October 2026, with fixed settings and weights, gives the AUC for such target earthquakes.

QuakeGPT learns only what the ETAS model contains. In the ETAS model, the rate of earthquakes became larger only after an earthquake occurred (Section 3.3). The months before new activity therefore contain no sign of the first earthquake of that activity. Without such a sign, $\Psi_{\mathrm{QuakeGPT}}$ does not decrease in the month in which new activity begins (Section 5.2).

## 6 Conclusions

1. Off Hokkaido and Tohoku (Figure 1), we computed $\Psi$ from the JMA earthquake catalogue ($M \geq 2.5$, 2013–2026). $\Psi$ is the weighted mean of 101 members $\Theta_j$, one for each value of $\lambda_{\mathrm{min}}$. $\Psi$ therefore needs no choice of one value of $\lambda_{\mathrm{min}}$, whereas the AUC of a single member changes with $\lambda_{\mathrm{min}}$ (Figure S3). Among the evaluated months (Section 3.2), $\Psi$ tended to be higher in the months followed within $T_W = 1$ year by a target earthquake ($M \geq 6.5$) than in the other months (Figures 2 and 3a). Random scores used in place of $\Psi$ did not show this tendency (Figure 3).
2. We trained QuakeGPT, a transformer, only on $\Psi$ of a 1000-year synthetic catalogue simulated with an ETAS model (Section 3.3). Years 901–1000 of this catalogue were used neither for training nor for selecting the QuakeGPT that was kept. $\Psi_{\mathrm{QuakeGPT}}$, for the next month, was almost the same as $\Psi$ in these years (Figure 6) and in the region (Figure 2). $\Psi_{\mathrm{QuakeGPT}}$ did not decrease in the month in which new activity began (Figures 2 and 6b).

3. Beyond the data, QuakeGPT gives a path of Ψ, with a range, for about one year (Figure 2). The path consists of values of Ψ, not of earthquakes. The present Ψ already concerns target earthquakes in the year after the last month of data (Section 5.2).
4. From earthquake monitoring, earthquake nowcasting with QuakeGPT can be used to update the present Ψ and the path of Ψ beyond the data (Figure 2). This set can therefore be used together with long-term assessments and with the information issued after large earthquakes. In this article, we proposed how nowcasting can contribute to the multi-hazard monitoring needed for earthquake preparedness.

## Acknowledgements

We thank the JMA for the earthquake catalogue. K.Z.N. carried out the early part of this study as a Visiting Faculty member at the Department of Physics and Astronomy, University of California, Davis, in 2025–2026. Research by K.Z.N. was partially supported by MEXT under the Second Earthquake and Volcano Hazards Observation and Research Program (Earthquake and Volcano Hazard Reduction Research) and the Earthquake Disaster Prevention Research Project to Minimize Damage from Gigantic Earthquakes such as the Nankai Trough Earthquake and Contribute to Rapid Recovery and Reconstruction, and by the SECOM Science and Technology Foundation (FY2023 Special Research Grant). Research by J.B.R. was supported in part by the Southern California Earthquake Center (grant SCON-00007927) and by the John LaBrecque Fund, University of California, Davis. Claude (Anthropic) was used under the direction of the authors. The authors reviewed all content and take full responsibility for it.

## Conflict of interest statement

The authors declare no conflicts of interest.

## Data availability statement

The JMA earthquake catalogue is provided to registered users (https://www.data.jma.go.jp/eqev/data/catalog/). The depth of the plate boundary, the upper surface of the Pacific plate, was obtained from https://www.mri-jma.go.jp/Dep/sei/fhirose/plate/en.index.html (Nakajima & Hasegawa, 2006; Kita et al., 2010). The Python code used in this study and the processed catalogues are available from the corresponding author upon reasonable request. The original QuakeGPT code is available at https://doi.org/10.5281/zenodo.11630546 (Rundle et al., 2025). XETAS is described by Ogata and Tsuruoka (2016). SASeis2006, on which it is based, is available at https://www.ism.ac.jp/editsec/csm/index_j.html.

## Figure captions

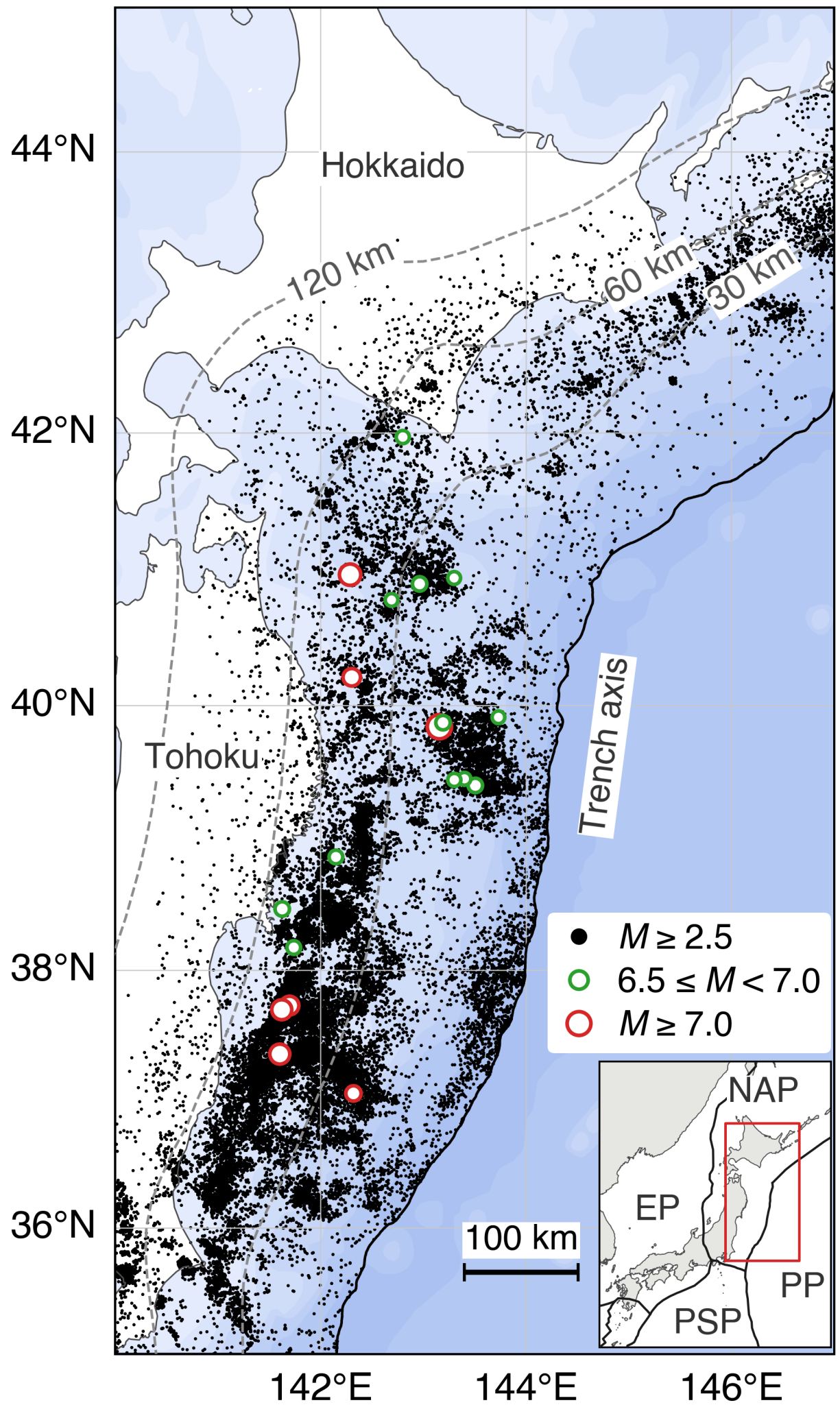

**Figure 1.** Study region off Hokkaido and Tohoku, Japan. Black dots indicate the earthquakes with $M \geq 2.5$ from January 2013 to October 2026 on and around the plate boundary, selected as described in Section 2. Circles are the 19 target earthquakes with $M \geq 6.5$ (green, $6.5 \leq M < 7.0$; red, $M \geq 7.0$). Larger circles show a larger $M$. The black line is the trench axis. Grey dashed lines show the depth of the plate boundary (30, 60, and 120 km). Blue shading shows the sea depth. The inset shows the location of the study region (red box) and the plate boundaries of Bird (2003). PP, Pacific plate; PSP, Philippine Sea plate; EP, Eurasian plate; NAP, North American plate. Map lines delineate study areas and do not necessarily depict accepted national boundaries.

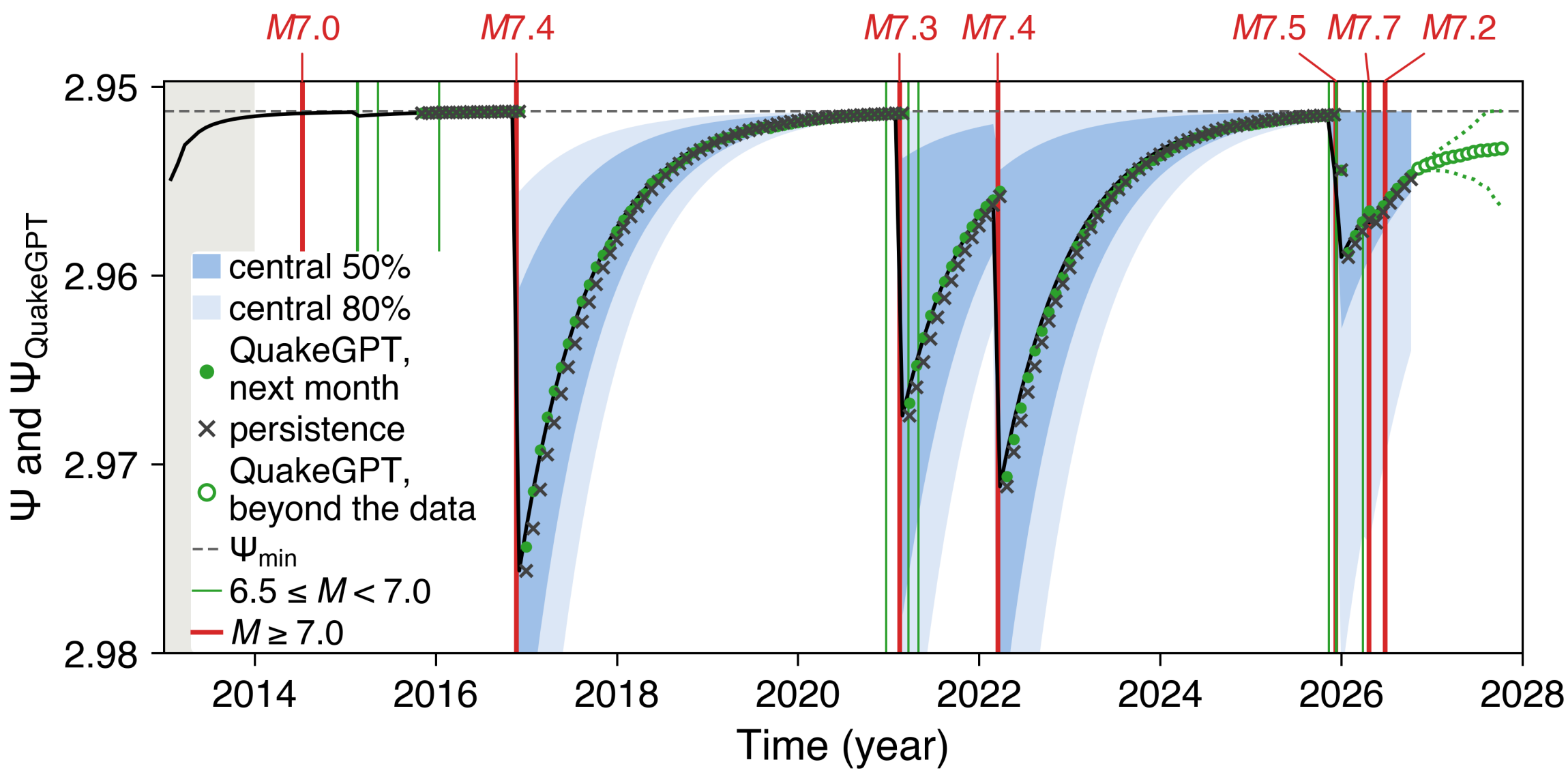

**Figure 2.** $\Psi$ and $\Psi_{\text{QuakeGPT}}$ as a function of time for the observed catalogue. The black line is $\Psi$, the weighted mean of the 101 members $\Theta_j$. The dashed horizontal line is $\Psi_{\min}$. The dark and light blue bands contain the central 50% and 80%, respectively, of the members shifted as in Section 3.2. The vertical axis is inverted, as in Rundle et al. (2025). The grey area marks 2013, the start-up year of the EMA, which was not evaluated. Vertical lines mark the target earthquakes (thin green, $6.5 \leq M < 7.0$; thick red, $M \geq 7.0$). The magnitudes of the $M \geq 7.0$ earthquakes are indicated above the frame. The green dots are $\Psi_{\text{QuakeGPT}}$ for the next month, computed from the $\Psi$ of the preceding 36 months, so that the first of these values is for November 2015. The crosses are persistence, that is, the $\Psi$ of the month before. The green open circles are $\Psi_{\text{QuakeGPT}}$ beyond the data for one year, that is, for the 13 months after the last month of data. The green dotted lines are these values plus and minus the median of $|\Psi - \Psi_{\text{QuakeGPT}}|$ (black line in Figure 7d), but not above $\Psi_{\min}$.

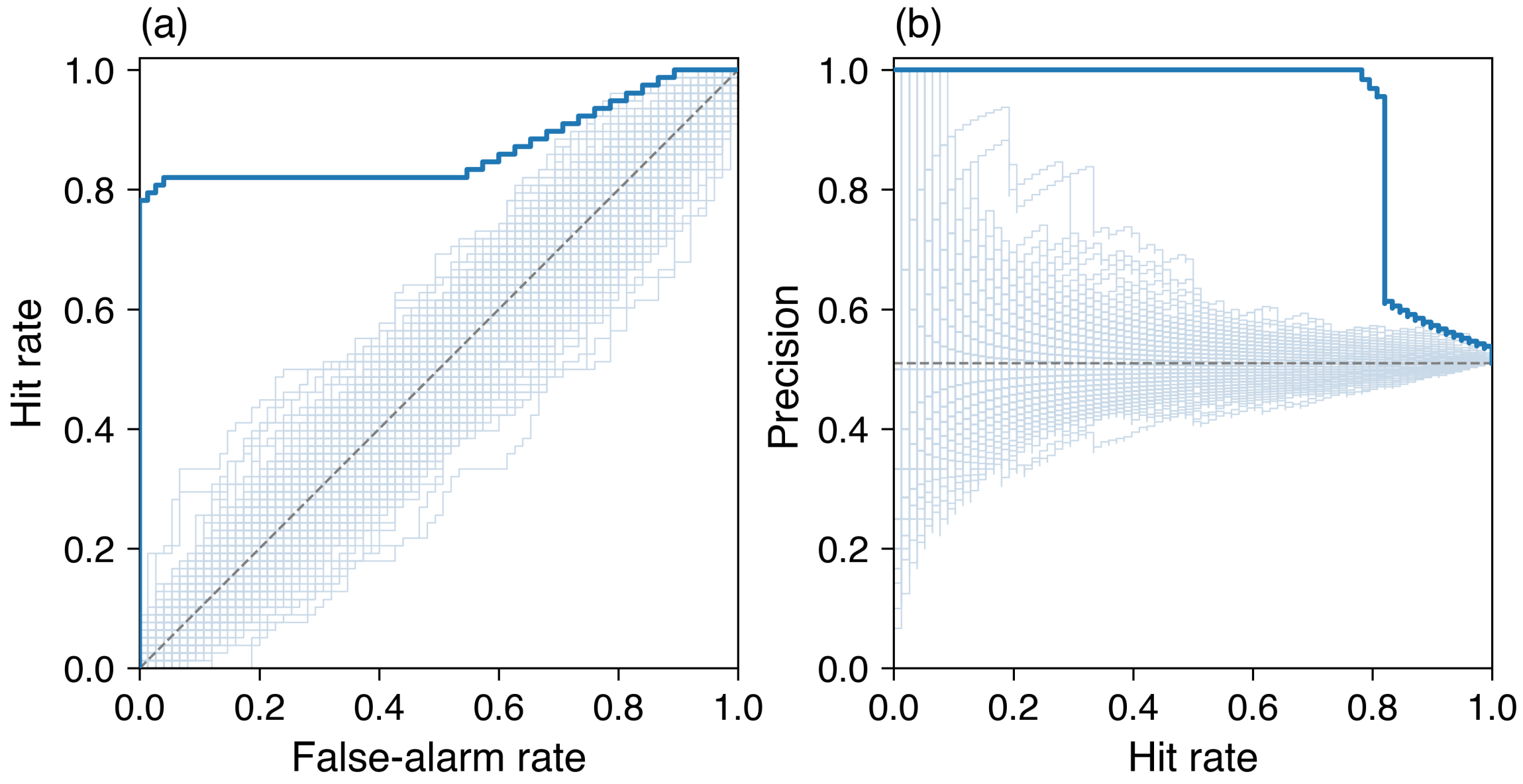

**Figure 3.** Skill of Ψ for $T_W = 1$ year (thick blue curve). (a) ROC curves. (b) Precision–recall curves. The dashed line is the level of no skill (AUC = 0.5 in a; the base rate, 0.51, in b). The light blue curves are those of 200 series of random scores.

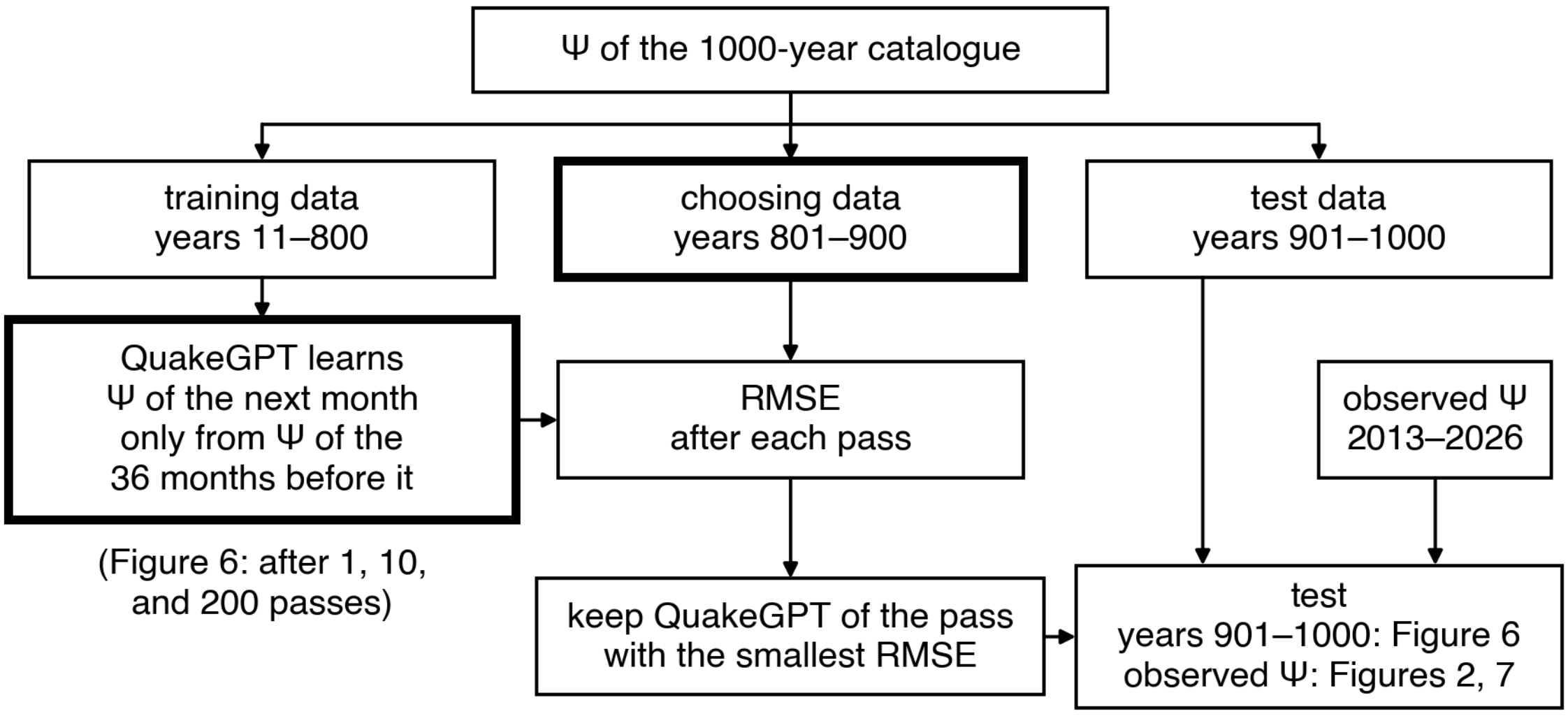

**Figure 4.** How QuakeGPT was trained and tested. Thick-lined boxes: the two changes to the code published by Rundle et al. (2025) that are needed for a valid test.

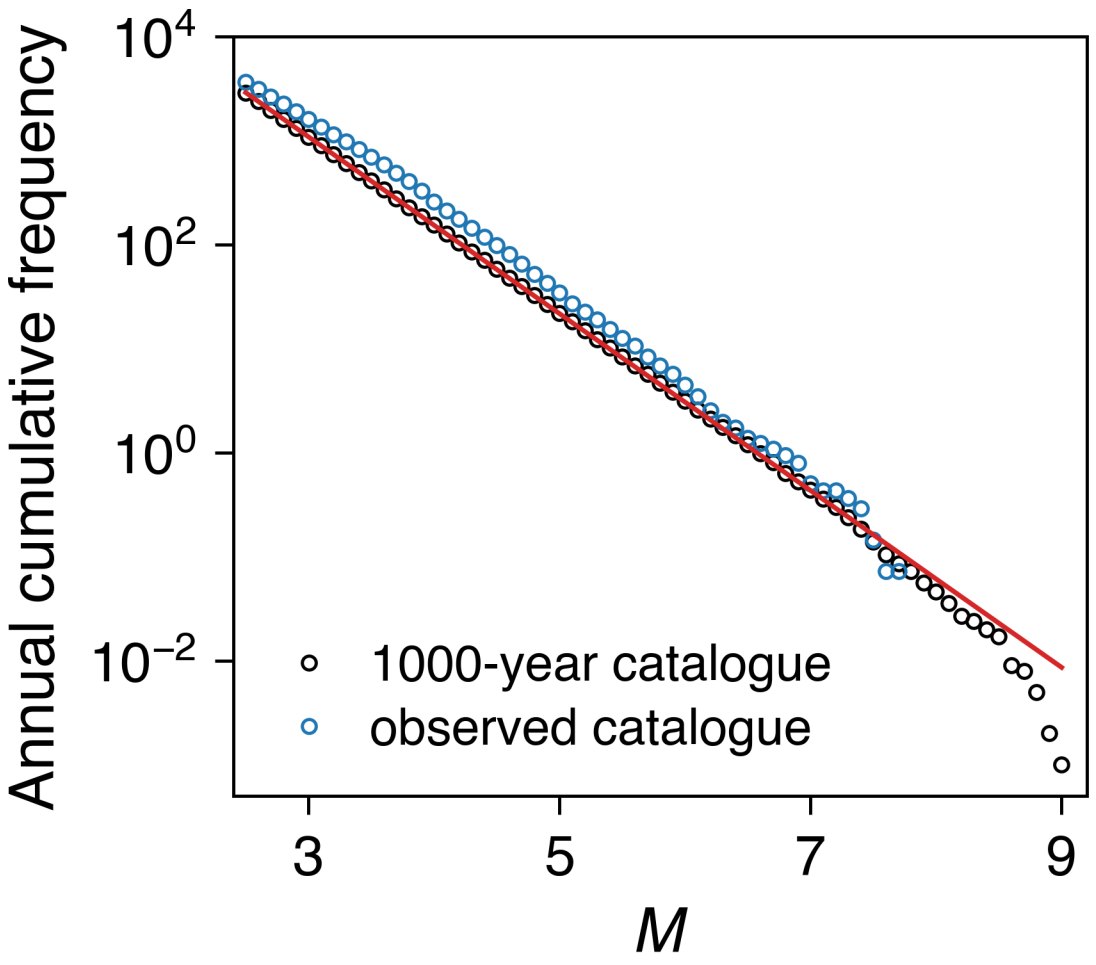

**Figure 5.** Annual cumulative frequency, that is, the number of earthquakes per year with magnitude *M* or larger. Black is the 1000-year catalogue, a synthetic catalogue from the ETAS model, and blue is the observed catalogue of January 2013 to October 2026. The red line is the Gutenberg–Richter relation with $b = 0.85$.

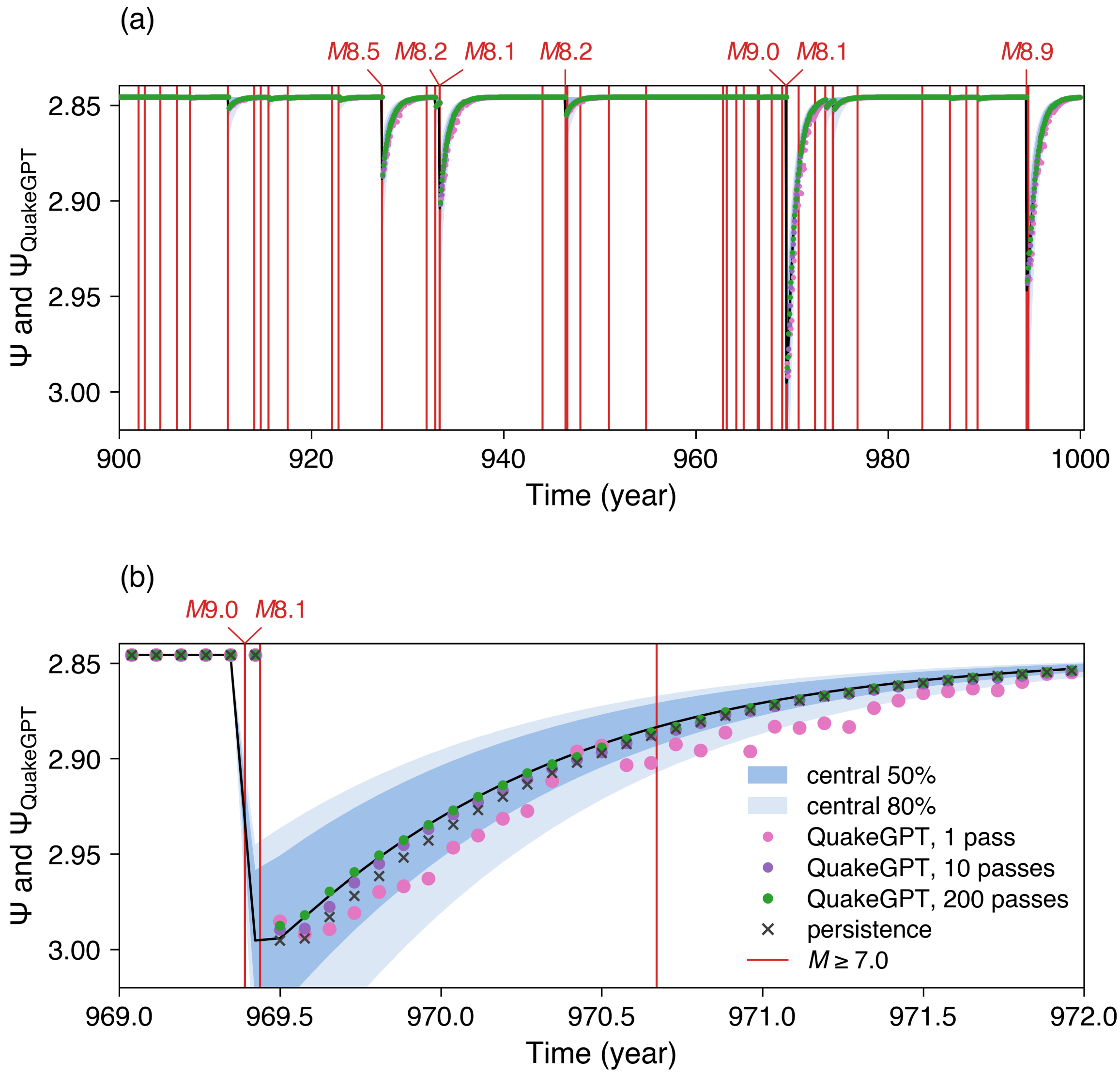

**Figure 6.** Test of QuakeGPT on the 1000-year catalogue. (a) Years 901–1000, the test data. (b) Years 970–972. For details of the vertical axis, the black line, the dark and light blue bands, and the crosses, see the caption of Figure 2. The dots are $\Psi_{\text{QuakeGPT}}$ for the next month for three lengths of the training: 1 pass (pink), 10 passes (purple), and 200 passes (green). Red vertical lines mark the earthquakes with $M \geq 7.0$, and the magnitudes of the $M \geq 8.0$ earthquakes are given above the frame.

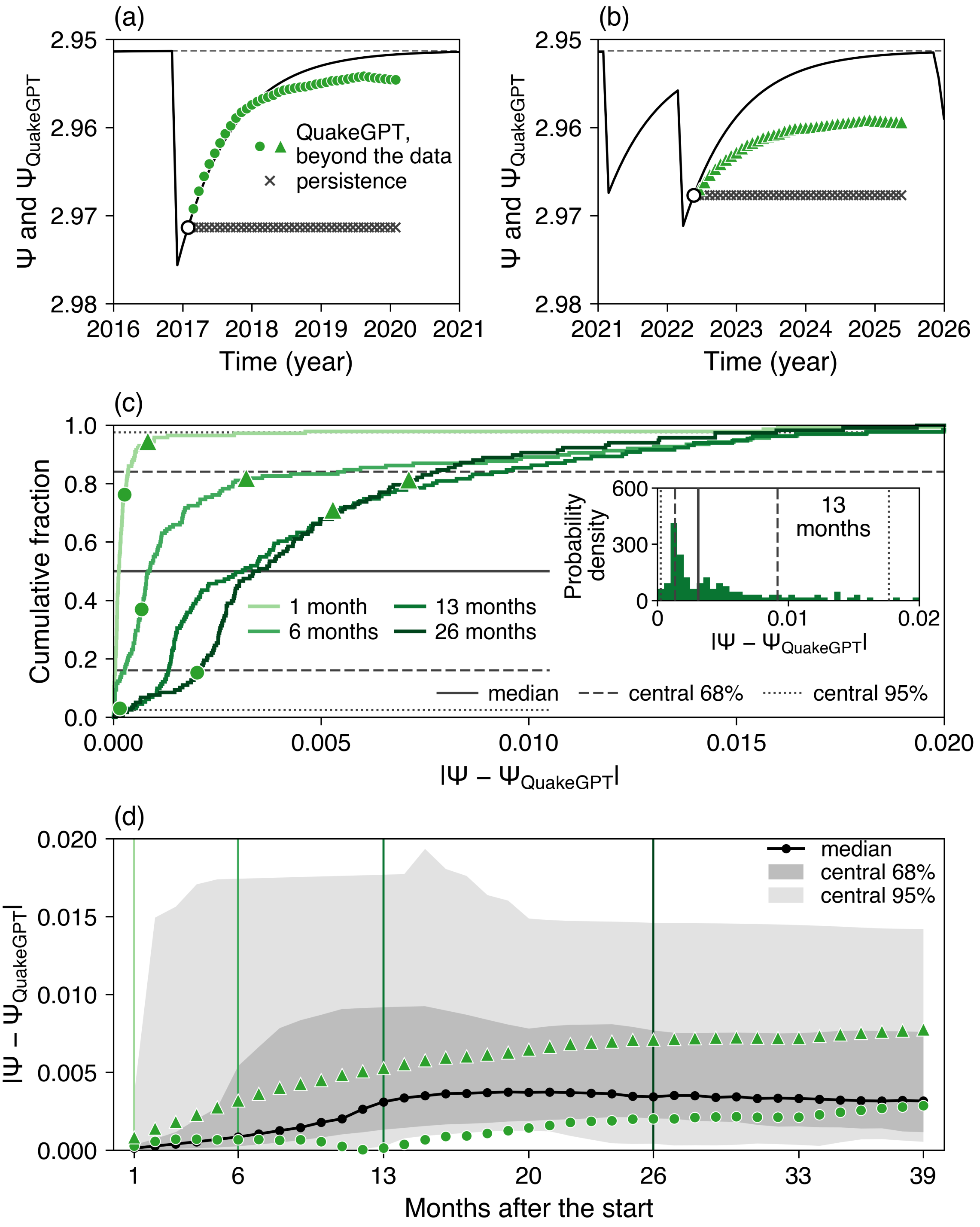

**Figure 7.** $\Psi_{\text{QuakeGPT}}$ beyond the data for the observed catalogue. (a, b) Two examples, started in (a) January 2017 and (b) May 2022. For details of the vertical axis, the black line, and the dashed line, see the caption of Figure 2. The black open circle is the $\Psi$ of the start month, the green dots (a) and triangles (b) are $\Psi_{\text{QuakeGPT}}$ for the 39 months after the start month, and the crosses are persistence. (c) Fraction of the start months for which $|\Psi - \Psi_{\text{QuakeGPT}}|$ is at most the value on the horizontal axis. Curves: 1, 6, 13, and 26 months after the start (light to dark green). Green dot and triangle on each curve: $|\Psi - \Psi_{QuakeGPT}|$ of (a) and (b), respectively. Horizontal lines at fractions of 0.5 (median, solid), 0.16 and 0.84 (central 68%, dashed), and 0.025 and 0.975 (central 95%, dotted). Inset: probability density of $|\Psi - \Psi_{QuakeGPT}|$ 13 months after the start, with its median, central 68%, and central 95% drawn as vertical lines. (d) Median (black line), central 68% (grey), and central 95% (light grey) of $|\Psi - \Psi_{QuakeGPT}|$ over all start months against the number of months after the start. Green dots and triangles: $|\Psi - \Psi_{QuakeGPT}|$ of (a) and (b), respectively. Vertical lines: 1, 6, 13, and 26 months after the start, in the colors of (c).

# Supplementary Material

# Nowcasting earthquakes with QuakeGPT off Hokkaido and Tohoku, Japan

K. Z. Nanjo, J. Yazbeck, I. T. Baughman, J. Donet, J. B. Rundle

## Text S1. EPS

Time is divided into months of equal length, $\delta = 0.07692$ year (one thirteenth of a year). The first month starts at 00:00 on 1 January 2013. The study period contains 179 months, numbered $k = 1, \ldots, 179$. Let $n_k$ be the number of earthquakes with $M \geq 2.5$ in month $k$. Let $m$ be the mean of $n_k$ over these 179 months. For this study period, $m = 282$. Months with few earthquakes are raised to the floor (Section 3.2), and the raised number of month $k$ is

$$x_k = \max(n_k, R_{\min}), \quad R_{\min} = \lambda_{\min}\, m \qquad \text{(S1)}$$

In the published code, the cap (Section 3.2) was $R_{\max} = R_{\min} + \lambda_{\max}\, s_n$, where $s_n$ is the standard deviation of $n_k$. The EMA of $x_k$ over the past months is

$$E_k = \frac{\sum_{h=0}^{k-1} u_h x_{k-h}}{\sum_{h=0}^{k-1} u_h}, \qquad u_h = (1-\gamma)^h, \qquad \gamma = \frac{2}{L+1} \qquad \text{(S2)}$$

where $h$ counts the months back from month $k$. The weight $u_h$ becomes smaller by the factor $1 - \gamma$ for each month further back. Rundle et al. (2025) wrote $N$ and $\alpha$ for $L$ and $\gamma$. We renamed them because $N$ and $\alpha$ denote other quantities in Text S4. The value of $\Theta$ in month $k$ is

$$\Theta_k = \log_{10}(1 + E_k) \qquad \text{(S3)}$$

In quiet months, $n_k$ does not exceed $R_{\min}$, so that $x_k = R_{\min}$ and $\Theta$ stays at the level set by the floor. When $n_k$ of a month exceeds $R_{\min}$, $E_k$ and the value of $\Theta$ become larger. On the inverted axis, this is a decrease of $\Theta$ (Section 3.1). If $n_k$ does not exceed $R_{\min}$ in the following months, the excess of $E_k$ over $R_{\min}$ is multiplied by about $1 - \gamma$ each month, and $\Theta$ returns to the level set by the floor. For $L = 20$ months (Section 3.2), the excess becomes 1/e (37%) of its initial value after about 10 months, 0.8 years.

Figure S1 shows the AUC of $\Theta$ against $L$ for $\lambda_{\min} = 2.5$ and 4.0. For each of these two values, we call a value of $L$ typical if AUC − 0.5, the amount by which the AUC exceeds the value without skill, is between 68% and 95% of its largest value over $L = 1$–1500 months. The typical values of $L$ form the grey areas in Figure S1. We used $L = 20$ months, which is typical for both values of $\lambda_{\min}$ (Section 4.2).

## Text S2. Ensemble EPS

From $\mathrm{AUC}_j$ (Text S3), we computed the z-score $z_j$ of member $j$, and from $z_j$ the weight $w_j$,

$$z_j = \frac{\mathrm{AUC}_j - \langle \mathrm{AUC} \rangle}{s_{\mathrm{AUC}}}, \qquad w_j = \frac{\max(0, z_j)}{\sum_{l=1}^{101} \max(0, z_l)} \tag{S4}$$

Here $\langle \mathrm{AUC} \rangle$ and $s_{\mathrm{AUC}}$ are the mean and the standard deviation of $\mathrm{AUC}_j$ over all 101 members, and $\Sigma_{j=1}^{101} w_j = 1$.

The value of $\Psi$ (Section 3.2) in month $k$ is $\Psi_k = \Sigma_{j=1}^{101} w_j \, \Theta_{j,k}$. Here $\Theta_{j,k}$ is the value of $\Theta_j$ in month $k$. Member $j$ has its own floor, $R_{\min,j}$ (Equation (S1)). When $n_k$ does not exceed $R_{\min,j}$ in any month, $\Theta_j$ equals the level set by its floor, $\log_{10}(1 + R_{\min,j})$. When all members are at these levels, $\Psi$ equals $\Psi_{\min} = \Sigma_{j=1}^{101} w_j \log_{10}(1 + R_{\min,j})$. The bands around $\Psi$ (Section 3.2) are computed with the weights $w_j$. Because the members are shifted, the bands show the spread of the time histories, and not the differences between the levels set by the floors (Figure S4a).

## Text S3. Evaluation of the EPS

We write $\tau_k$ for the end of month $k$. If at least one target earthquake occurs in the interval $(\tau_k, \tau_k + T_W]$, month $k$ is a positive month (Section 3.2). We evaluated $k = 14$ to 166 (Section 3.2).

The AUC of Ψ equals the probability that a randomly chosen positive month has a smaller Ψ than a randomly chosen negative month, with ties counted as one half. This probability is the Mann–Whitney U statistic divided by the number of pairs of a positive and a negative month (Mann & Whitney, 1947; Hanley & McNeil, 1982). We computed the AUC in this form, AUC = $P(\Psi_{pos} < \Psi_{neg}) + 0.5\, P(\Psi_{pos} = \Psi_{neg})$. Here, $\Psi_{pos}$ and $\Psi_{neg}$ are the values of Ψ in a positive and negative month, respectively, and $P$ is the probability over all such pairs. For Θ, the AUC is $P(\Theta_{pos} < \Theta_{neg}) + 0.5\, P(\Theta_{pos} = \Theta_{neg})$. Here, $\Theta_{pos}$ and $\Theta_{neg}$ are the values of Θ in a positive and negative month, respectively.

In each of the 200 series of random scores (Section 3.2; Figure 3), Ψ of each evaluated month was replaced by a random number, which we call the random score. Positive months stay positive, and negative months stay negative.

## Text S4. Synthetic catalogue from an ETAS model

We produced the 1000-year catalogue (Section 3.3) by simulation with the temporal ETAS model (Ogata, 1988). In this model, the number of earthquakes with $M \geq 2.5$ per day at time $t$, $\lambda(t)$, is

$$\lambda(t) = \mu + \sum_{i:t_i<t} K_0 \exp[\alpha(M_i - M_{\mathrm{ref}})]\,(t - t_i + c)^{-p} \qquad \text{(S5)}$$

Here, μ is the background rate, that is, the rate of earthquakes that are not aftershocks. The sum runs over all earlier earthquakes $i$. Each term of the sum is the rate of aftershocks of earthquake $i$. The time of earthquake $i$ is $t_i$, and its magnitude is $M_i$. The aftershock rate of an earthquake decays as an inverse power of the time that has passed since it, $t - t_i$, with the constants $c$ and $p$. A larger earthquake has more aftershocks. This dependence on $M_i$ is expressed by $\exp[\alpha(M_i - M_{\mathrm{ref}})]$, with the constant α. Here, $M_{\mathrm{ref}}$ is a reference magnitude. For an earthquake with $M_i = M_{\mathrm{ref}}$, $\exp[\alpha(M_i - M_{\mathrm{ref}})]$ equals 1, so that $K_0$ sets the number of its aftershocks. In view of the 2011 Tohoku earthquake (*M*9.0), it is reasonable to assume that $M$9.0 is the largest magnitude in this region. We therefore used $M_{ref} = 9.0$, the same value as $M_{max}$ (Section 3.3).

We estimated μ, $K_0$, $c$, α, and $p$ by the maximum-likelihood method (Ogata, 1988) from the earthquakes with $M \geq 2.5$ between 1 January 2023 and 8 October 2026 in the study region and east of it up to 148°E (11,818 earthquakes). We used the program package XETAS (Ogata & Tsuruoka, 2016), which adds a graphical user interface to the program package SASeis2006 (Ogata, 2006). The estimates and their standard errors are $\mu = 4.52 \pm 0.14$ per day, $K_0 = 291.5 \pm 24.2$, $c = 0.0134 \pm 0.0016$ day, $\alpha = 1.64 \pm 0.03$, and $p = 1.087 \pm 0.013$. The period was limited to the years from 2023, because the program only accepts a limited number of earthquakes.

Equation (S5) gives the rate at which earthquakes occur, but not their $M$. In the simulation, $M$ of each earthquake was drawn at random from the Gutenberg–Richter relation (Gutenberg & Richter, 1944), $\log_{10} N = a - bM$, with $b = 0.85$ (Section 3.3). Here, $N$ is the number of earthquakes of magnitude $M$ or larger, and $a$ is a constant. The simulated earthquakes have $M$ from 2.5 (Section 2) to $M_{max} = 9.0$.

In most months of the 1000-year catalogue, $n_k$ is about 200 (Figure S5). In the months of the largest earthquakes, $n_k$ exceeds 1000. In the month of the $M$9.0 earthquake, $n_k$ is about 3600.

For Ψ of the 1000-year catalogue, $m$ is the mean of $n_k$ over years 1–900, so that the test data play no part in the training or in the choice of the QuakeGPT that was kept.

## Text S5. QuakeGPT

QuakeGPT computes its output from its input in a long series of simple operations. These operations contain adjustable numbers, which are called internal parameters. With the structure of the published code (Section 3.3), QuakeGPT has 35,777 internal parameters.

The rescaling of each input window (Section 3.3) is linear. After the rescaling,

QuakeGPT receives neither the values of Ψ nor the size of their changes. The rescaled value computed by QuakeGPT was converted back to $\Psi_{QuakeGPT}$ with the smallest and largest values of the input window.

During training, dropout was applied (Srivastava et al., 2014). Dropout means that, each time the internal parameters are adjusted, a randomly chosen part of QuakeGPT (20%, as in the published code) is switched off. QuakeGPT must then work well, even when a part of it is missing. The purpose of dropout is to keep QuakeGPT from simply memorizing the training data.

Training for too long can make $\Psi_{QuakeGPT}$ close to the true values in the training data but less close in other months (Srivastava et al., 2014). For this reason, the QuakeGPT that was kept was chosen on the choosing data (Section 3.3; Figure 4). QuakeGPT computes Ψ of each month of the choosing data as for the next month. For each of the three lengths of the training in Figure 6 (1, 10, and 200 passes), the QuakeGPT of the pass with the smallest RMSE (Section 3.3) among the passes made was kept. The tests on the observed Ψ use the QuakeGPT kept after 200 passes.

We made five changes to the published code. The first two are described in Section 3.3. The other three changes are described next. Third, dropout was used only while the internal parameters were adjusted. QuakeGPT computes Ψ without dropout, both for the choosing data and in the tests. In the published code, dropout remained active when Ψ of the next month was computed, so that $\Psi_{QuakeGPT}$ of a month changed at random from one computation to the next. Fourth, the true value of the next month was rescaled with the smallest and largest values of its own input window and not, as in the published code, with those of the whole time series of Ψ. The true value thus has the same scale as the input window. Fifth, the output of QuakeGPT is the change of Ψ from the last month of the input window to the next month, and this change is added to the last value. In the published code, the output was the value itself. With the change of Ψ as the output, QuakeGPT had to learn only the departures from no change. We selected this output because Ψ changed little in most months (Figures 2 and 6).

Beyond the data (Section 3.3), Ψ of the first month after the start month was computed as for the next month. To compute Ψ of the second month, the input window was moved forward by one month. The input window then consisted of the last 35 true values and $\Psi_{\text{QuakeGPT}}$ of the first month. This step was repeated month by month.

## Supplementary References

## Supplementary Figures

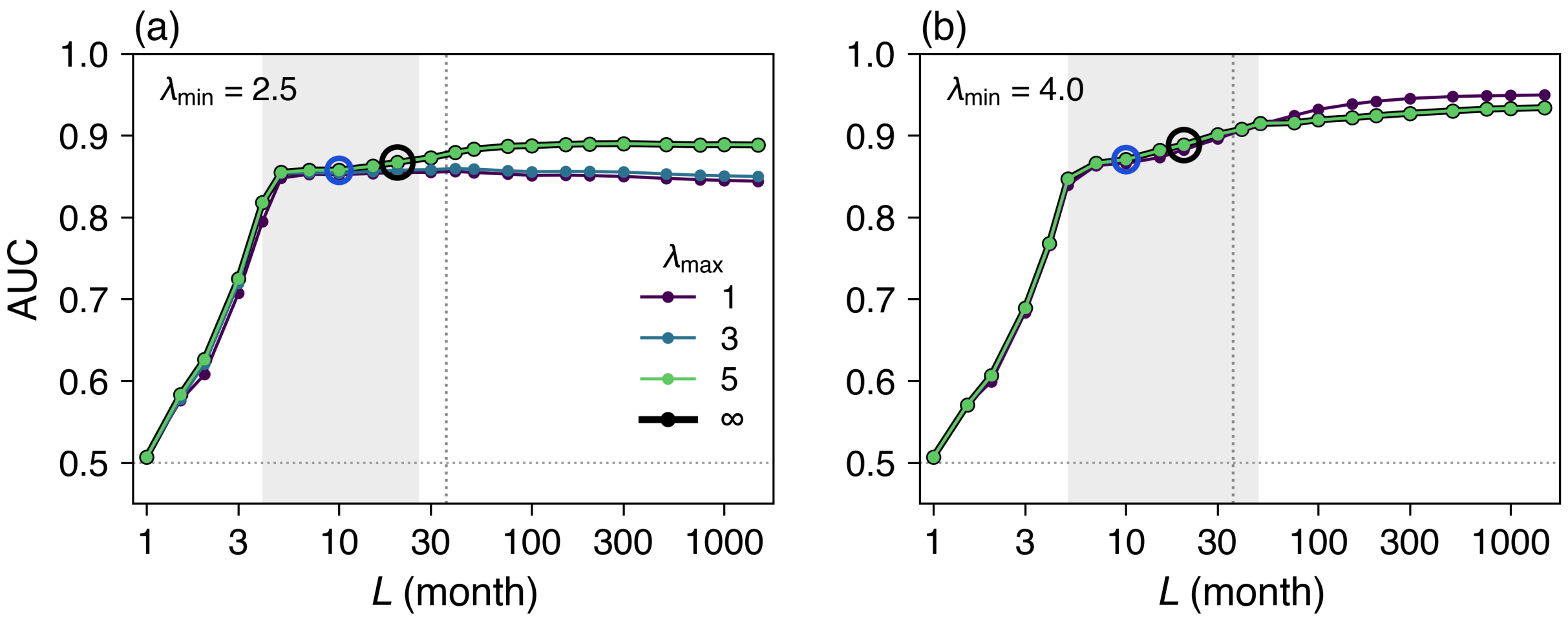

**Figure S1.** Choice of $L$ (Text S1). The panels show the AUC of $\Theta$ against $L$, evaluated as in Text S3, for (a) $\lambda_{min} = 2.5$ and (b) $\lambda_{min} = 4.0$. The coloured lines are for $\lambda_{max} = 1$, 3, and 5 (Text S1). The black line, drawn beneath them, is for $\lambda_{max} = \infty$, as used in this study. In (a), the line for $\lambda_{max} = 5$ lies on the black line; in (b), the lines for $\lambda_{max} = 3$ and 5 do. The grey area marks the typical values of $L$ defined in Text S1 (68–95% of the rise of the AUC from 0.5 to its maximum). The circles mark $L = 20$ months (black), used in this study, and $L = 10$ months (blue), used to repeat the analysis of the EPS (Figure S2). The vertical dotted line marks $L =$ 36 months, used by Rundle et al. (2025). The horizontal dotted line is the level of no skill (AUC = 0.5).

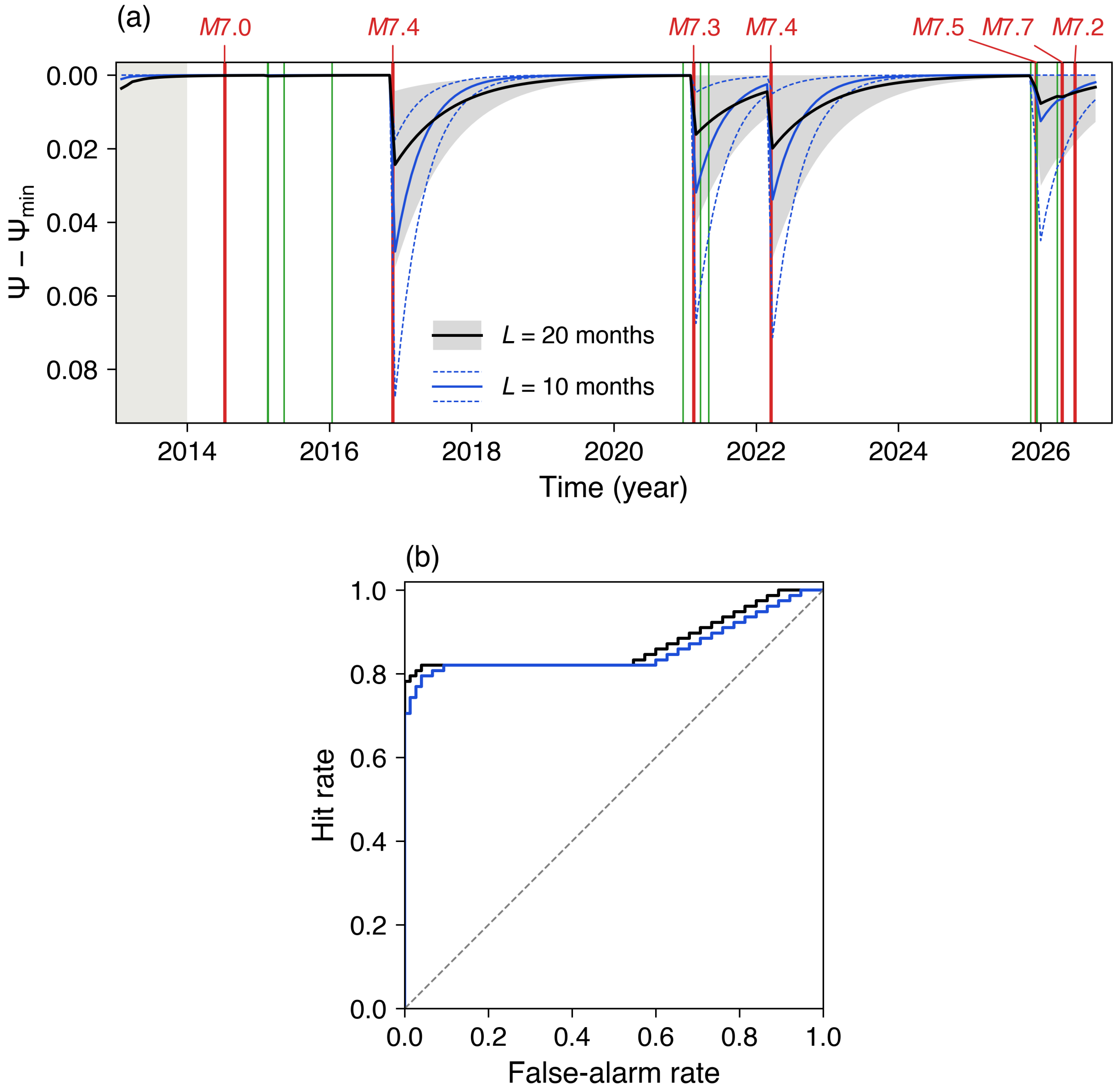

**Figure S2.** Analysis of the EPS repeated with $L$ = 10 months (Section 4.2). Each panel adds the result for $L$ = 10 months (blue) to a figure of the main text for $L$ = 20 months (black): (a) Figure 2 and (b) Figure 3a. (a) $\Psi - \Psi_{\mathrm{min}}$ (Text S2), with the $\Psi_{\mathrm{min}}$ of each $L$. The vertical axis is inverted, as in Figure 2. The central 80% of the members is shown, as in the light band of Figure 2 (grey band for $L$ = 20 months, dashed lines for $L$ = 10 months). The vertical lines and the grey area (2013) are as in Figure 2. (b) ROC curves, as in Figure 3a, but without the curves of random scores. The dashed line is the level of no skill.

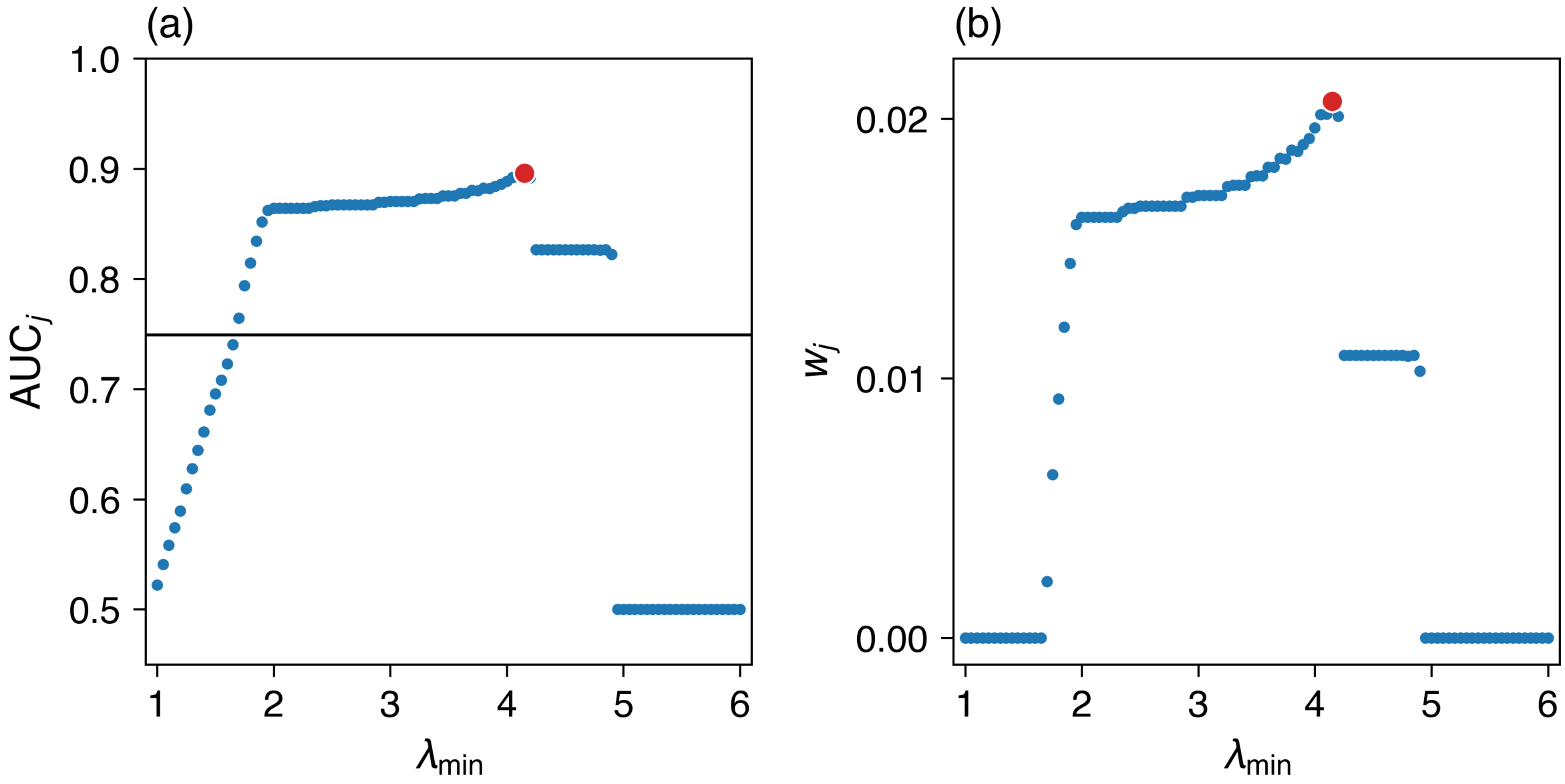

**Figure S3.** AUC and weights of the 101 members, against $\lambda_{min}$ (Text S2). Each blue dot is one member. (a) $AUC_j$. The horizontal line is the mean, $\langle AUC \rangle = 0.75$. (b) The z-score weight $w_j$ of Equation (S4). The red dots in (a) and (b) mark the member with the largest $AUC_j$, $\lambda_{min} = 4.15$. This member is shown in Figure S4b.

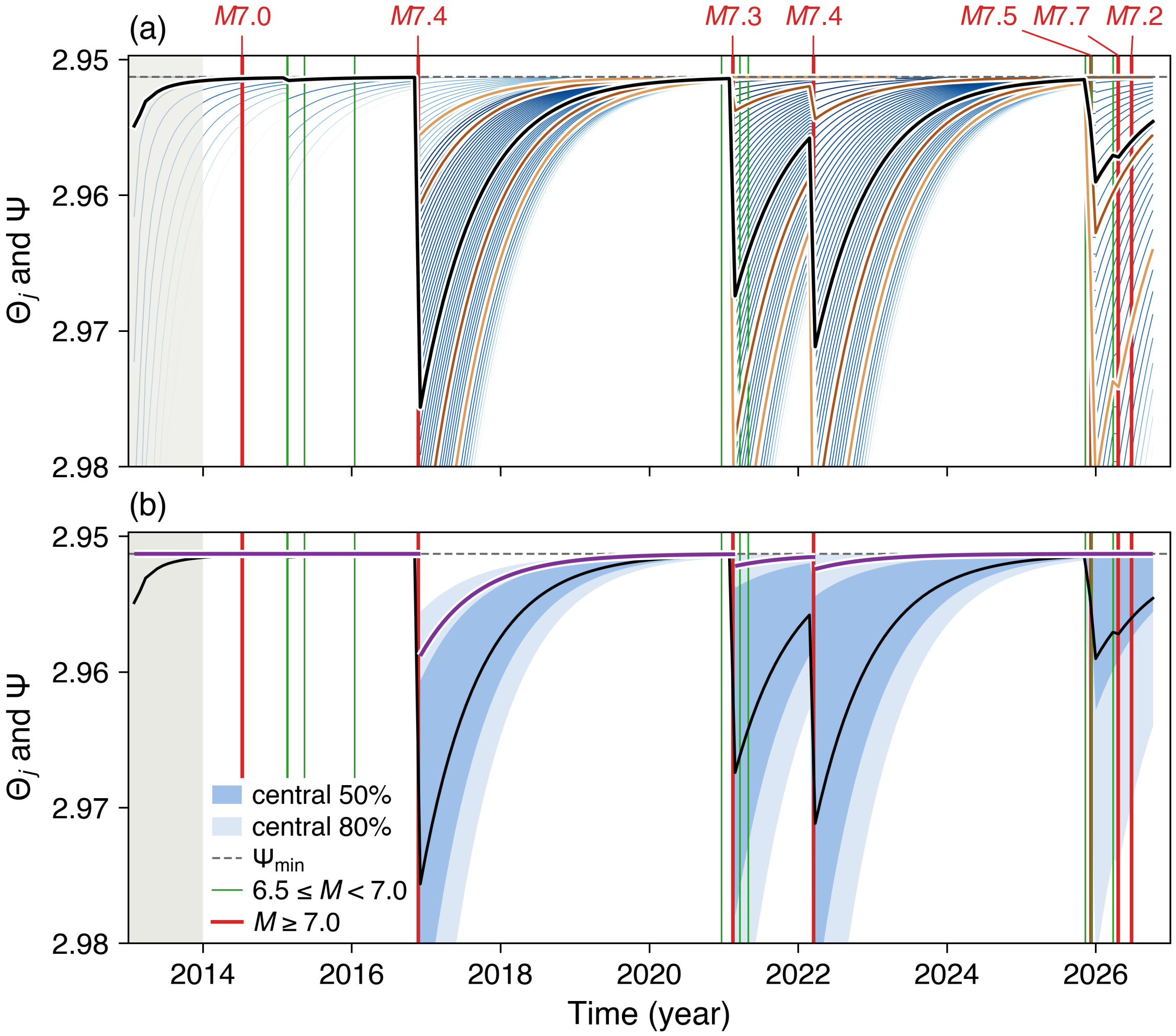

**Figure S4.** Members $\Theta_j$, shifted as in Section 3.2, with $\Psi$ as in Figure 2. (a) The 65 members $\Theta_j$ with $w_j > 0$ (blue; thicker and darker for larger $w_j$). The bands of Figure 2 are shown by their edges (dark brown, central 50%; light brown, central 80%). (b) $\Psi$ and its bands as in Figure 2, with a purple curve added. This curve is the member with the largest $AUC_j$, $\lambda_{min}$ = 4.15 (red dot in Figure S3).

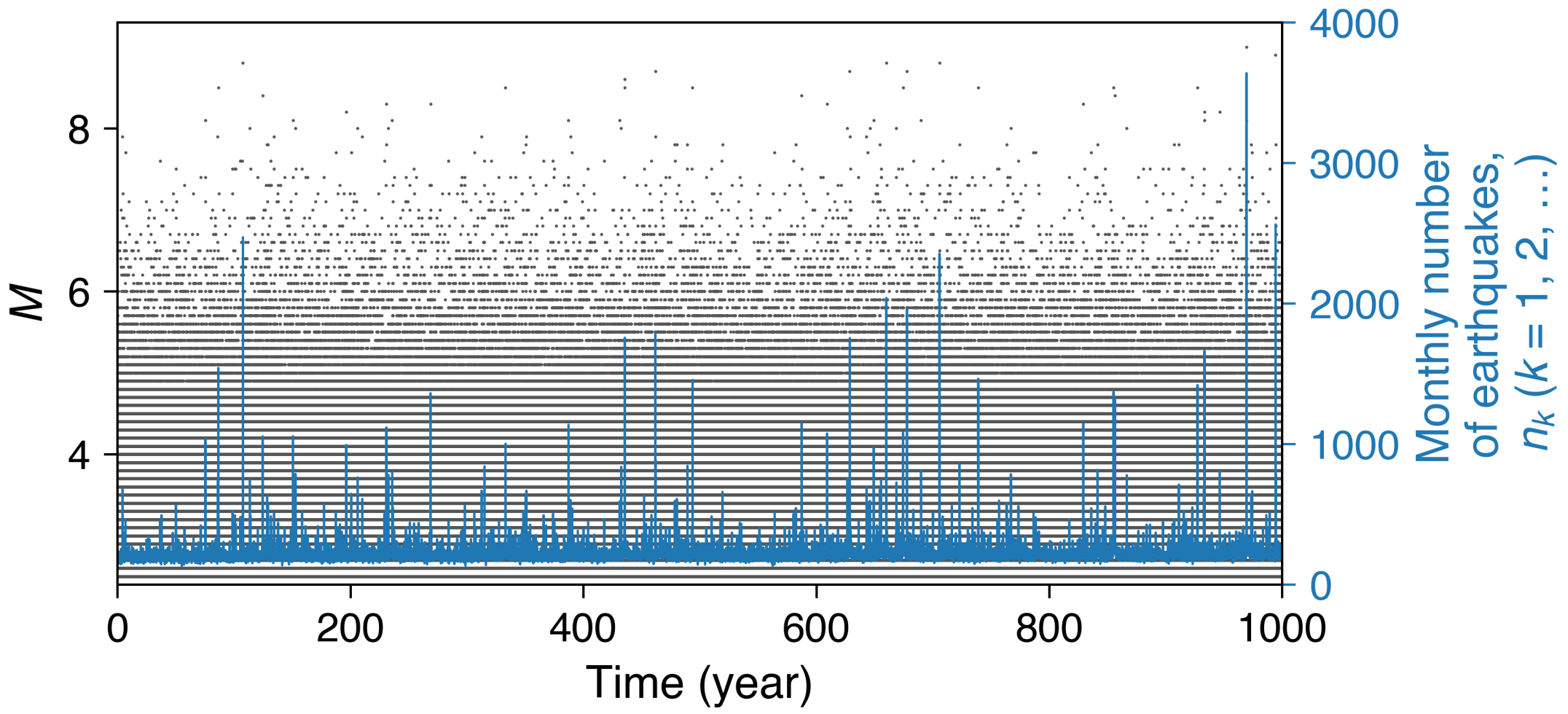

**Figure S5.** The 1000-year catalogue (Text S4). Grey dots are $M$ of all earthquakes ($M \geq 2.5$) against time (left axis). The blue line is $n_k$ ($k = 1, 2, \ldots$) (right axis).

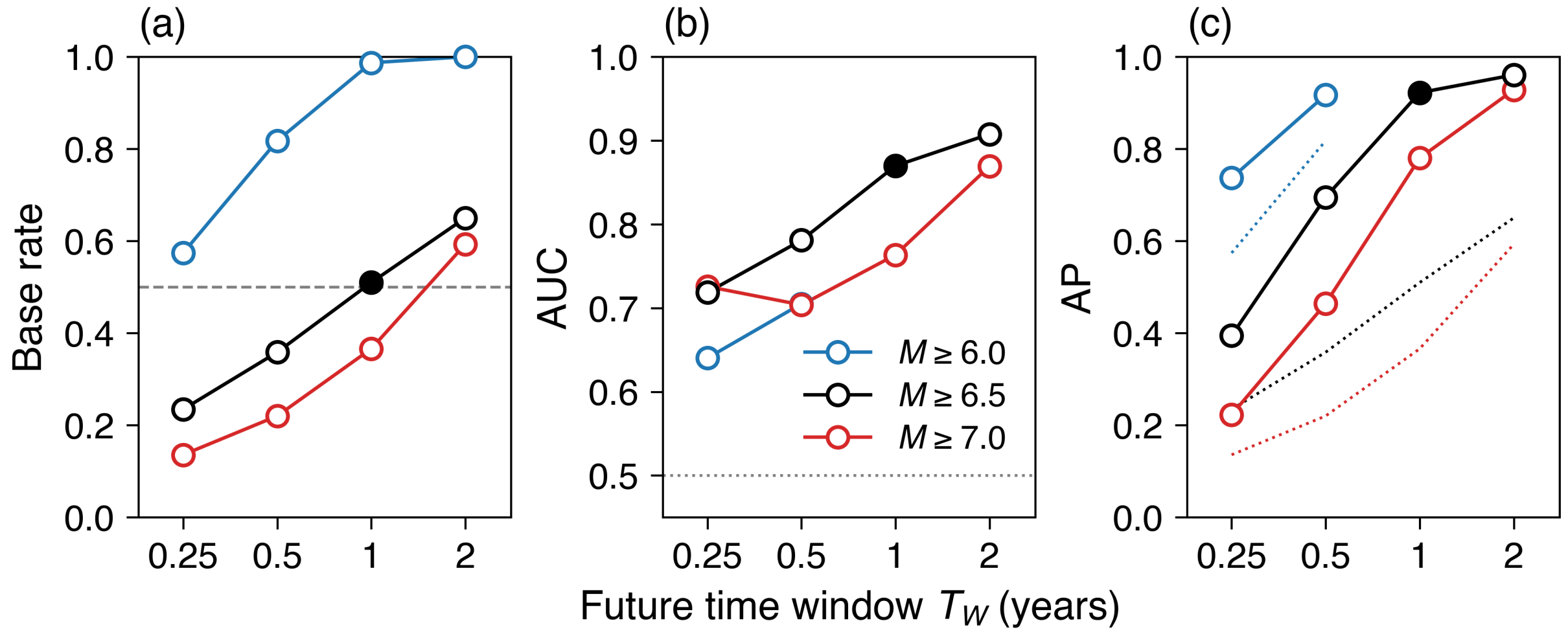

**Figure S6.** Skill of Ψ against $T_W$. Black is for the targets of the main analysis, $M \geq 6.5$. The filled circle is the main analysis ($T_W = 1$ year). Blue is for targets with $M \geq 6.0$, and red is for targets with $M \geq 7.0$. The weights are set for each case. The numbers of evaluated months are 162, 159, 153, and 140 for $T_W = 0.25$, 0.5, 1, and 2 years. For $M \geq 6.0$ and $T_W$ of 1 and 2 years, at most two months are negative, and AUC and AP are not given. The quantities are defined in Sections 3.1 and 3.2. (a) Base rate. The dashed line marks 0.5, at which positive and negative months are equal in number. (b) AUC. The dotted line is the level of no skill, 0.5. (c) AP. The dotted lines are its level of no skill, the base rate of (a), in the colour of the targets. Where circles overlap, black is drawn over red and red over blue (in (b) at $T_W = 0.25$ and 0.5 years, and in (c) at $T_W = 2$ years).